\documentclass[conference]{IEEEtran}
\IEEEoverridecommandlockouts

\usepackage{cite}
\usepackage{amsmath,amssymb,amsfonts}
\usepackage{algorithmic}
\usepackage{graphicx}
\usepackage{textcomp}
\usepackage{xcolor}
\usepackage[normalem]{ulem}
\graphicspath{ {images/} }
\usepackage{hyperref}

\usepackage{subcaption}
\def\BibTeX{{\rm B\kern-.05em{\sc i\kern-.025em b}\kern-.08em
    T\kern-.1667em\lower.7ex\hbox{E}\kern-.125emX}}
\begin{document}

\title{Scalability and Performance Evaluation of IEEE 802.11ah IoT Deployments: A Testbed Approach\\
}

\author{\IEEEauthorblockN{Kostas Chounos, Katerina Kyriakou, Thanasis Korakis}
\IEEEauthorblockA{\textit{Department of Electrical and Computer Engineering} \\
\textit{University of Thessaly, Volos, Greece}\\
\{hounos, aikyriakou, korakis\}\@uth.gr}
}

\maketitle

\begin{abstract}

This work focuses on the development and assessment of modern wireless Internet of Things (IoT) architectures, with relevance to emerging 5G and beyond applications. To analyze the growing demands for data, and their impact, we built an IEEE 802.11ah (WiFi Halow) office testbed for real-world experimentation. This deployment allows us to uncover the practical performance and scalability limitations of such networks under various challenging scenarios. To the best of our knowledge, this is the first study to consider complex real-world IEEE 802.11ah implementations, aiming specifically to reveal unexpected performance behaviors, such as significant throughput degradation arising in closely deployed wireless links. Our findings show that intense network contention and Adjacent Channel Interference (ACI), drastically impact the performance of the wireless links involved. Beyond evaluating network performance, our experimental analysis also considers the energy consumption of the devices under test, offering a more holistic perspective on the feasibility of IEEE 802.11ah in real-world deployments. The effective disclosure of such unexpected phenomena, can lead to well planned decisions and energy consumption optimization across the IoT to Cloud continuum.

\end{abstract}




\section{Introduction} 

During the past decades, the world witnessed an exponential demand \cite{CISCO} for constantly connected users to the Internet. As a result, the sharply increased number of interconnected devices in modern networks, dragged along the user data demands. Thus, several significant challenges have emerged such as network \& energy efficiency, management, security, interoperability and scalability, which are intensively examined from the research community in recent years. Despite that all facts mentioned above are crucial for the networks' performance, it is also widely acknowledged that interference \cite{Interference_IoT}  tends to be one of the most intensive problems for the wireless networks. The vast amount of data generated by numerous devices places a strain on existing infrastructures, often resulting in increased latency and degraded performance due to packet collisions or extended back-off times. 

\begin{figure*}[t!]
    \begin{center}
      	\includegraphics[width=0.99\linewidth]{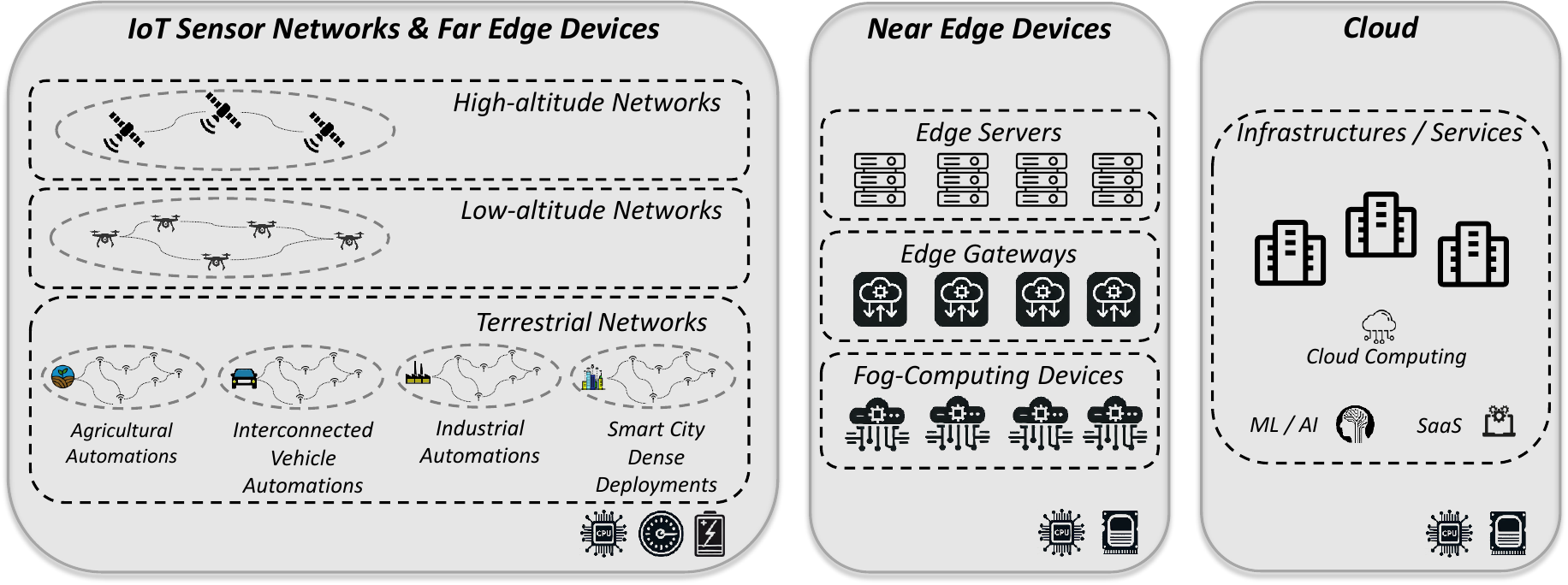}
    	\caption{5G \& beyond, IoT Things to Cloud Continuum.}
        \label{fig:IoT_continuum}
    	\vspace{-0.5cm}
    \end{center}
\end{figure*}

The next generation of wireless networks, particularly within the 6G \cite{IoT_6G_general} era, are expected to support a wide range of critical IoT-driven applications that demand massive data processing, Machine Learning (ML) integration, and seamless connectivity across diverse environments. Key application scenarios include but are not limited to Unmanned Aerial Vehicle (UAV) operations, industrial automation, and Autonomous Vehicle to everything (V2X) communications \cite{CSCN23_UAV, IoT_Ind_Automation, IoE_autonomous_vehicles}, all of which will be fundamental to future smart cities. These use cases include several deployment types, ranging from sparse to ultra-dense networks, with a crucial aspect of these deployments to be the IoT-to-cloud continuum as depicted in Figure \ref{fig:IoT_continuum}. There, real-time data processing, edge computing, and cloud integration play a vital role in ensuring efficiency and scalability. The challenge lies not only in handling large volumes of data but also in optimizing energy consumption across all the parts of the continuum. Energy-efficient communication, adaptive resource allocation, and intelligent data offloading mechanisms are essential to sustaining the viability of IoT networks in such data-intensive applications. Addressing these challenges is critical for ensuring the strong participation of IoT ecosystems in 6G networks.

Thus, there is an imperative need to develop efficient wireless protocols, which will be able to keep up both with the increased data demands and interference levels given in the modern IoT application scenarios. IEEE 802.11ah \cite{IEEE_802.11ah} (Wi-Fi HaLow) can be considered as a propitious solution to some of the problems mentioned above. This protocol supports from short $(<2m)$ up to long ranges $(<=1km)$ \cite{Practical_halow} with low power consumption profiles, thus covering a large number of the upcoming 6G IoT use cases. Compared to other widely adopted IoT protocols like LoRaWAN \cite{LoRaWAN} and NB-IoT \cite{NB-IoT}, it offers significantly larger achieved data rates, by utilizing the same sub-GHz unlicensed bands. The performance of this protocol is thoroughly examined and analyzed in Section \ref{evaluation}, by also taking into consideration existing limitations given at the commercial hardware, and the current regulations applied at each country.

The context of this work is to experimentally examine the behavior of this protocol under various challenging scenarios, and by building an indoor office testbed. Therefore, the realistic performance of IEEE 802.11ah is being observed when intense interference or contention is present in closely deployed devices. This work aims to reveal non-expected spectrum behaviors which are caused by ACI, and highlight how these strongly affect the expected network performance. The lower performance that is unexpectedly observed, directly affects the energy consumption of the involved devices and can be considered to extract more appropriate networking or processing decisions, along all parts of the IoT continuum. 

The remainder of this work can be summarized as follows. Initially, Section \ref{related} refers to the available approaches that can be found in the literature and are correlated with this work. Additionally, Section \ref{deployed_testbed} describes the testbed's topology, as well as the hardware and software components that were utilized for building the experimental IEEE 802.11ah wireless infrastructure. Furthermore, Section \ref{evaluation} contains a description of the experimental scenarios executed and the performance results obtained. Finally, Section \ref{conclusions} concludes the work and presents possible extensions as future work.

\section{Related Work} \label{related}

There are already published research articles in the literature, that aim to note the significant impact of interference in IoT dense architectures. These works include both generic approaches for heterogeneous wireless technologies operating in the same frequencies \cite{IoT_interference_NB-IoT, IoT_interference_general1, IoT_interference_general2}, and interference management when devices of the same protocol coexist \cite{IoT_interference_halow}. These approaches largely differ to each other, based on the operating frequencies of the involved IoT devices, as there are cases of utilization for both the licensed and the unlicensed spectrum. In the context of this work, we will focus on a developed testbed deployment for IEEE 802.11ah devices, which operate in Europe's (EU) unlicensed 868MHz (863–870 MHz Sub-GHz) band.

The highly unpredictable conditions prevail in the unlicensed spectrum environments, include several additional challenges in the channel access techniques and in the interference management as well. Thus, there are available research approaches proposed lately, which focus on predicting the interference \cite{IoT_unlicensed_interference_prediction, TNSE23_interference} levels, and detecting spectrum irregularities \cite{CSCN23_ML} through ML, within the limited available unlicensed spectrum. It should be noted that IEEE 802.11ah considers wireless contention and interference management, with built-in Medium Access Control (MAC) mechanisms, like Restricted Access Window (RAW) \cite{RAW}, or the widely applied Clear Channel Assessment (CCA) \cite{CCA}. However, there is a significant lack of practical approaches in the literature to demonstrate the proper functioning of the developed mechanisms, as most studies regarding IEEE 802.11ah are based solely on simulations. \cite{IEEE802.11ah_ns-3_2}. Although this is a standard established in 2017, commercial hardware has been rolled-out over the past two years. Until that point, minimal experimental approaches were available \cite{IEEE802.11ah_USRP}, which utilized dedicated devices like the Universal Software Radio Peripheral (USRP) to implement the IEEE 802.11ah stack in real radios. Along these lines, the scale up of the examined experimental topologies was nearly impossible to be achieved.  

On the other hand, interesting findings have been revealed through some early experimental approaches of the IEEE 802.11ah standard. Initially, the authors in \cite{Practical_halow} evaluate the aforementioned wireless protocol in both indoor laboratory and outdoor campus environments, combining Line of Sight (LoS) and NLoS cases between the nodes. More specifically, they follow a single Access Point (AP) - multiple Station (STA) approach focusing solely on wireless contention, and they compare the experimental performance findings to the theoretically values expected. The key metrics that they take into consideration are the Received Signal Strength Indicator (RSSI), Modulation Coding Scheme (MCS), Round Trip Time (RTT) of the transmitted packets, when compared to the data rates achieved. The results in some of the experimental scenarios tested, seem to keep up sufficiently with the theoretically ones expected. However, this is not the case in \cite{CSCN23_ML, IEEE802.11ah_experimental_Japan}, where the authors of both works note significant performance losses for the IEEE 802.11ah protocol, even in relatively simple topologies with up to four nodes, including APs and/or STAs. In the former work, the authors examined the practical performance of IEEE 802.11ah for Japan's sub-GHz (921-927 MHz) band, with 1, 2 and 4 MHz wide channels. Several physical (PHY) and MAC configurations were tested, and results demonstrated significant data transmission limitations for both indoor and long-range (emulated with attenuators) use cases. In the latter study, performance irregularities are observed even in relatively simple network architectures. To address this, the authors propose a centralized offline coordination algorithm that demonstrates improvements within only the specific set-up evaluated.    

The contributions of this work can be summarized as follows. Initially, we build an indoor office wireless infrastructure for experimenting with dense IEEE 802.11ah topologies, which are foreseen to have a significant involvement in the upcoming network architectures. This infrastructure provides both reproducible experimental results as it operates under controlled spectrum environment, and easy access / management to the experimenter. Furthermore, through extensive experimentation, the realistic performance and the overall behavior of these environments are being observed, under the presence of both heavy ACI and contention. Thus, this work aims to reveal and highlight multiple indicative problematic scenarios that cause significantly lower performance than expected, that will be taken into consideration to be resolved in real-time as a part of future work. Finally, the energy consumption for the developed nodes have been measured as well, which translates into additional energy costs, when severe performance issues occur.  

\section{Developed Infrastructure} \label{deployed_testbed}

\begin{figure*}[t!]
    \centering
    \begin{subfigure}{.5\textwidth}
        \centering
        \includegraphics[width=1\linewidth]{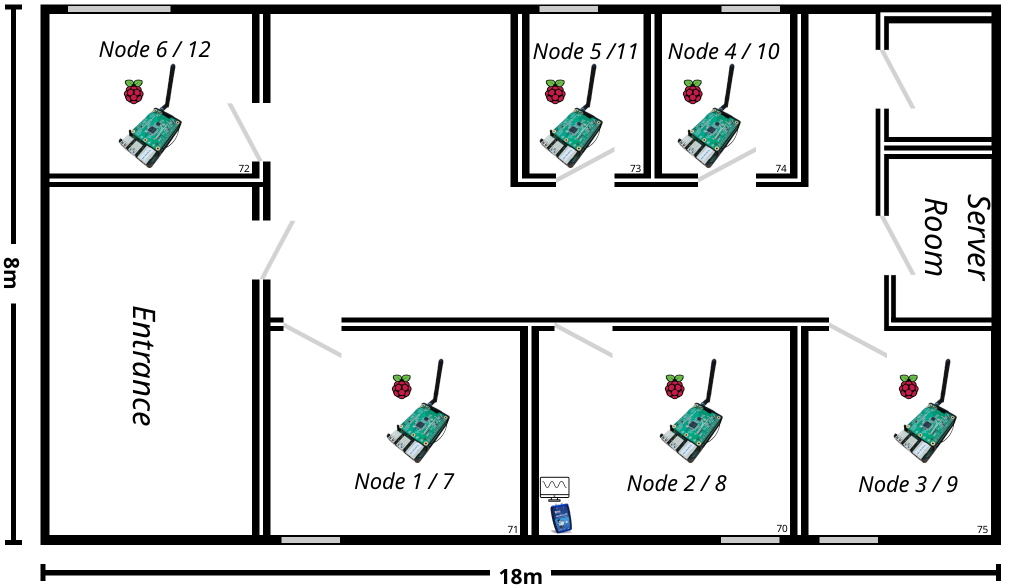}
    \end{subfigure}%
    \hspace{0.1cm}
    \begin{subfigure}{.45\textwidth}
        \centering
        \includegraphics[width=1\linewidth]{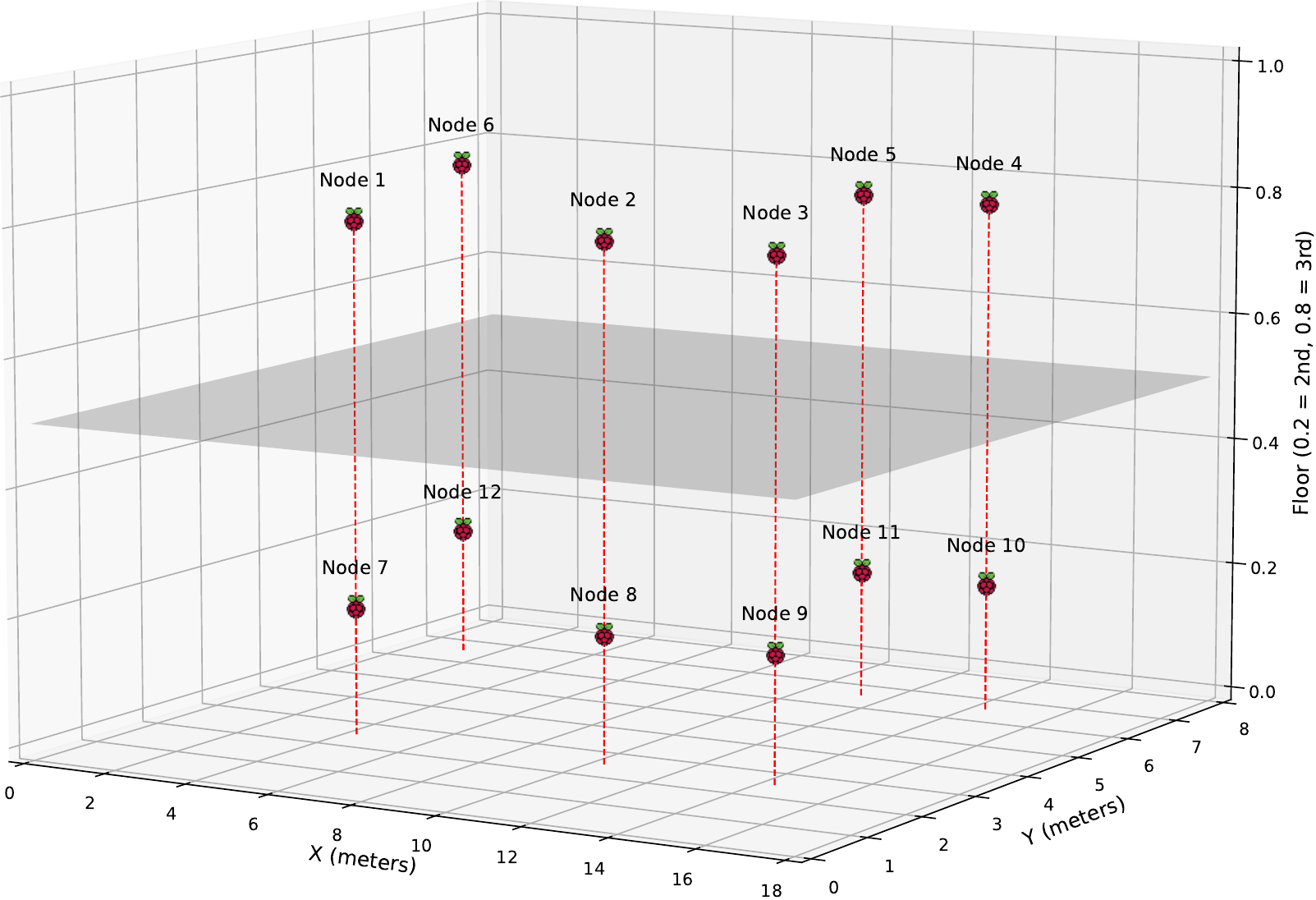}
    \end{subfigure}
    \caption{Testbed topology (Node X / Y for 3$^{rd}$ / 2$^{nd}$ floors accordingly).}
    \label{fig:testbed_topology}
    \vspace{-0.45cm}
\end{figure*}

In the context of this work, an IEEE 802.11ah indoor office testbed is developed and utilized to empower wireless experimentation in modern IoT environments. This section describes the topology, as well as all the software and hardware components involved in the developed infrastructure.

\subsection{Topology Description}

Initially, the infrastructure contains in its current state, twelve IEEE 802.11ah nodes placed in two different floors (2$^{nd}$ and 3$^{rd}$) of our research laboratory. 
Every floor is approximately $140m^2$, and each node is far from the others within a range of distances from $D_{min} = 2.5m$ to $D_{max} = 18m$. It is worth noting that both the placement of the nodes, as well as the architectural design of both floors, are nearly identical, thus maximizing fair conditions and reproducible experimental scenarios. The topology for the nodes is depicted in Figure \ref{fig:testbed_topology}, which contains Nodes 1-6 in the 3$^{rd}$, and Nodes 7-12 in the 2$^{nd}$ floor accordingly. In addition, the internal walls between the offices are made of plasterboard and a concrete ceiling is given between the two floors. Finally, the nodes are placed in a height of approximately $1m$ (at desks), and it is noted that there are no other devices in the building, which can cause external interference, and thus performance alteration to the under-study scenarios. 

\subsection{Hardware / Software Components} \label{hardware_comp}

The main component for each testbed node is a Raspberry Pi 3B+ / 4 single-board computer, which acts as a host for the wireless adapters and the main processing unit. More specifically, this hardware contains an ARM-Cortex (A53 / A72) processor with 2-8 GB of RAM depending on the version. Each node runs on the conventional Raspbian Operating System (OS), which is officially supported and maintained by the manufacturer. Every testbed node also contains an ALFA AHPI7292S Raspberry Hat \cite{ALFA} with the first commercial NRC7292 IEEE 802.11ah \cite{NRC7292} wireless chipset. An omni-directional antenna of $3dBi$ gain is attached in the ALFA Raspberry IEEE 802.11ah Hats. Bearing in mind the current IEEE 802.11ah standard's regulations for EU \cite{Wlan_channels}, there are only 1MHz wide channels available for deployment, as 2MHz is now an obsolete bandwidth. In such way, we follow the regulations which are also applied in the latest NRC wireless drivers (v1.5.2), and we take under consideration only 1MHz channels in the evaluation section. The expected data rates \cite{sun2013ieee} in the 1MHz channels in MCS 7 range from 3 Mbits/sec with Long Guard Interval (GI) to 3.34 Mbits/sec with Short GI. In our practical evaluation, we use Long GIs and the highest MCS observed was 7, achieving in practice up to 2.4Mbps of beneficial throughput. 

Furthermore, an external spectrum analyzer has been implemented, by utilizing a Pluto Software Defined Radio (SDR) \cite{Pluto} and Python programming, which deployed near Node 2 (3$^{rd}$ floor). Hence, the prevailing spectrum conditions for both floors can be effectively monitored in real-time during the experiments. Thus, we are able to verify that the performance noted is not affected from external unauthorized transmissions. The deployed spectrum analyzer has been set to sense frequencies from 862.5MHz to 868.5MHz, thus covering all channels under evaluation. The results are refreshing every 0.1 second and they are plotted in a "waterfall" graph. The spectrum analyzer was utilized during the execution of all the experiments, but only some indicative snapshots are included in this work to further confirm and explain the performance results obtained.  

 Finally, tailored  Bash \& Python scripts developed in order to store several useful network metrics (like selected MCS, RSSI success / transmission rate etc.) during the experiments, which are exported either from Raspbian OS or the NRC7292 driver. It is worth to mention that, the saturaded UDP network traffic is generated / noted with iperf tool \cite{Iperf} and stored in CSV files, from the involved nodes in all experimental uses cases and scenarios. Finally, each node is remotely accessed through the widely known SSH network protocol.
\section{Experimental Results} \label{evaluation}

This section includes a detailed description for the scenarios, the results and the observations which were exported after extensive experimentation in the developed IEEE 802.11ah deployment. As mentioned also in the Section \ref{hardware_comp}, based on the IEEE 802.11ah EU standard regulations, five 1MHz channels are available and utilized in the context of this work. From now on we will refer to these channels as $Ch_{1}$ = 863.5 MHz, $Ch_{2}$ = 864.5 MHz, , $Ch_{3}$ = 865.5 MHz, , $Ch_{4}$ = 866.5 MHz, and $Ch_{5}$ = 867.5 MHz. It is worth noting that all these channels are non-overlapping to each other and in theory, no significant interference or performance degradation is expected between them. However, this is not the case in real-world experiments, which is significantly opposed to already conducted simulations. This is thoroughly proved and analytically described in the sub-sections given below. 

In prior to every experimental execution, several actions were taken in order to ensure that the involved links operate as expected. More specifically for each pair of nodes, it is ensured that the wireless link quality is stable. This value is being exported from the "iwconfig" wireless tool, which classifies link quality in a scale 0 to 70. This is directly correlated with the amount of data (MCS automatically selected from rate adaptation mechanism) that can be transmitted successfully at each link. Before running any experimental scenario, a throughput test under completely clear channel conditions, was executed for each link in order to note the throughput achieved. This is in practice, the maximum throughput capacity which can be achieved, and it is noted as \textbf{\textit{"Link Capacity without Interference"}} in the figures below. For all scenarios executed in this work, saturated (above link's maximum capabilities) UDP traffic from STA to AP was generated with the iperf tool. Finally, the experimental duration were selected to be 120sec, and each scenario was executed 5 times in order to extract coherent results. Continuous monitoring through the deployed spectrum analyzer was also carried out during all the experimental scenarios execution, ensuring that the results were not affected by external uncoordinated factors. 


\begin{figure*}[t!]
    \centering
    \begin{subfigure}{.47\textwidth}
        \centering
        \includegraphics[width=1\linewidth]{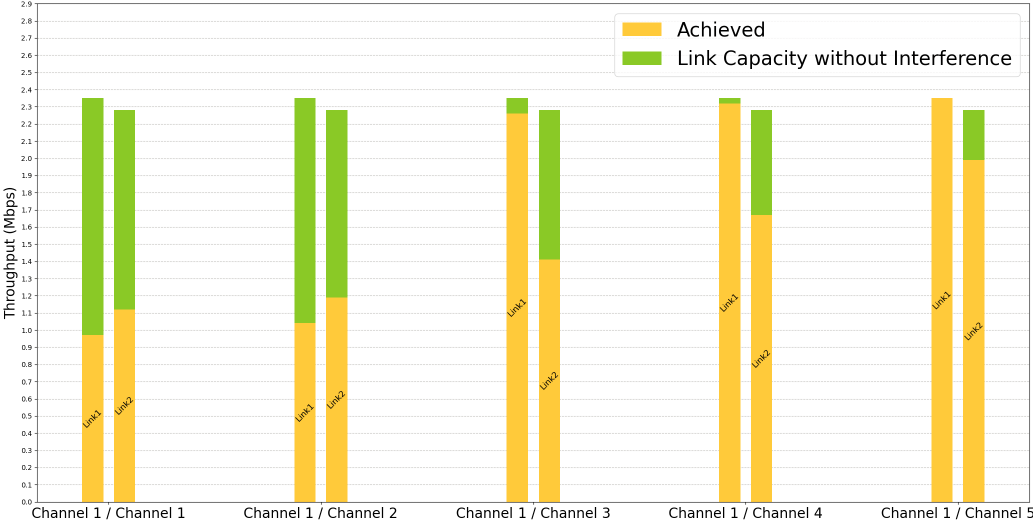}
        \caption{Throughput}
        \label{fig:ACI_2links_throughput}
        \end{subfigure}%
        \begin{subfigure}{.475\textwidth}
        \centering
        \includegraphics[width=1\linewidth]{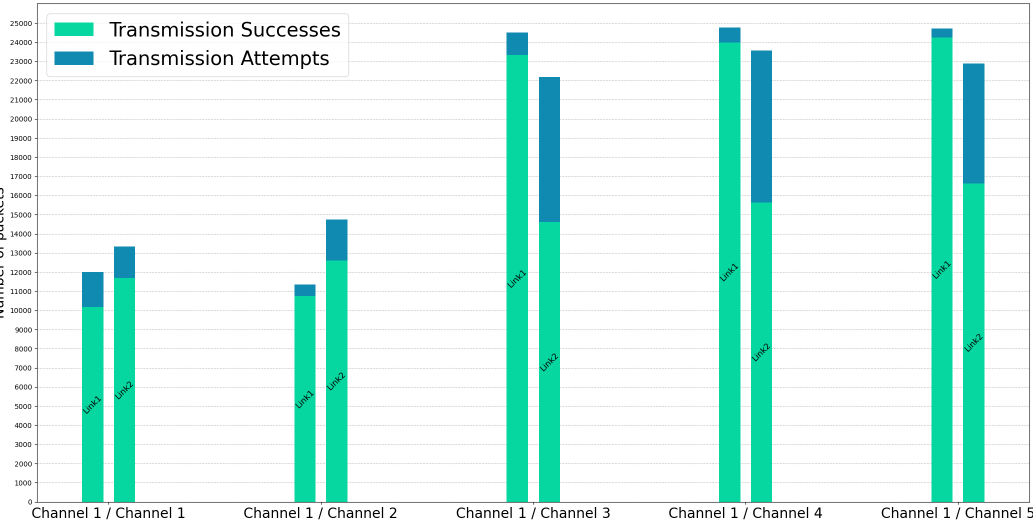}
        \caption{Packet Successes / Attempts}
        \label{fig:ACI_2links_packets}
    \end{subfigure}
    
    \vspace*{.2cm}
    
    \begin{subfigure}{.48\textwidth}
        \centering
        \includegraphics[width=1\linewidth]{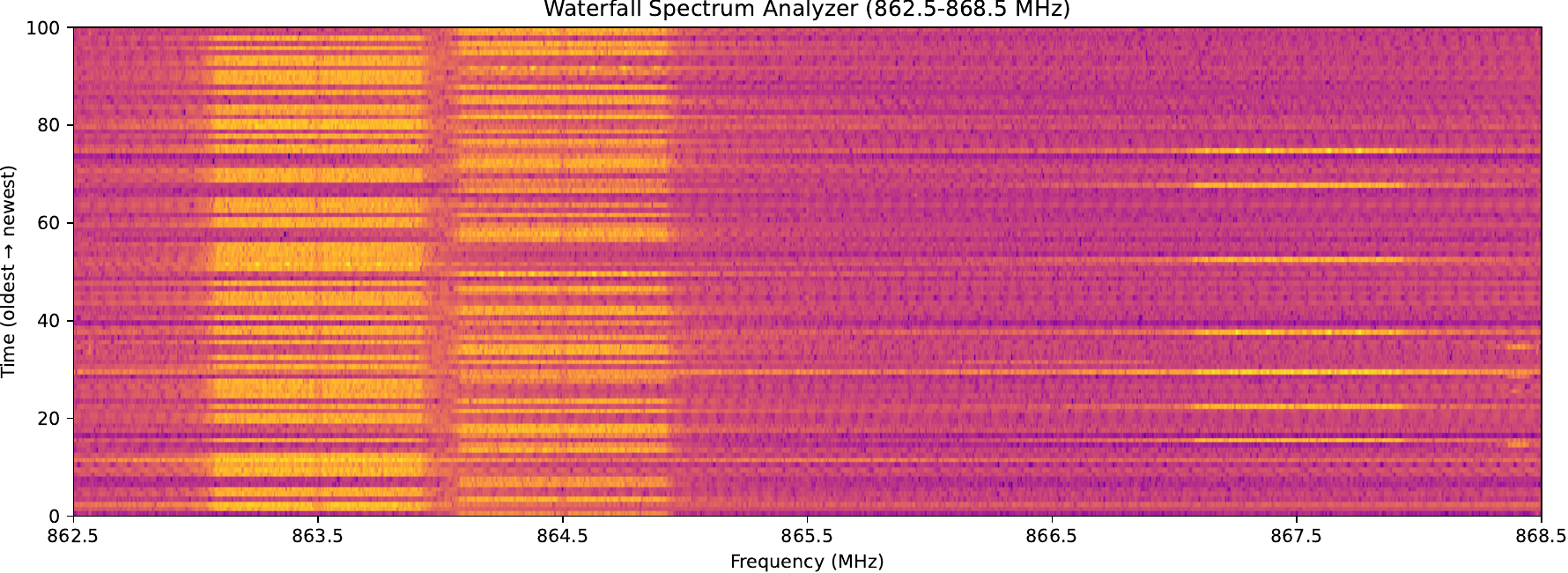}
        \caption{Spectrum Analyzer (Sub-scenario 2)}
        \label{fig:spectrum_analyzer_aci_1_2}
        \end{subfigure}%
        \begin{subfigure}{.48\textwidth}
        \centering
        \includegraphics[width=1\linewidth]{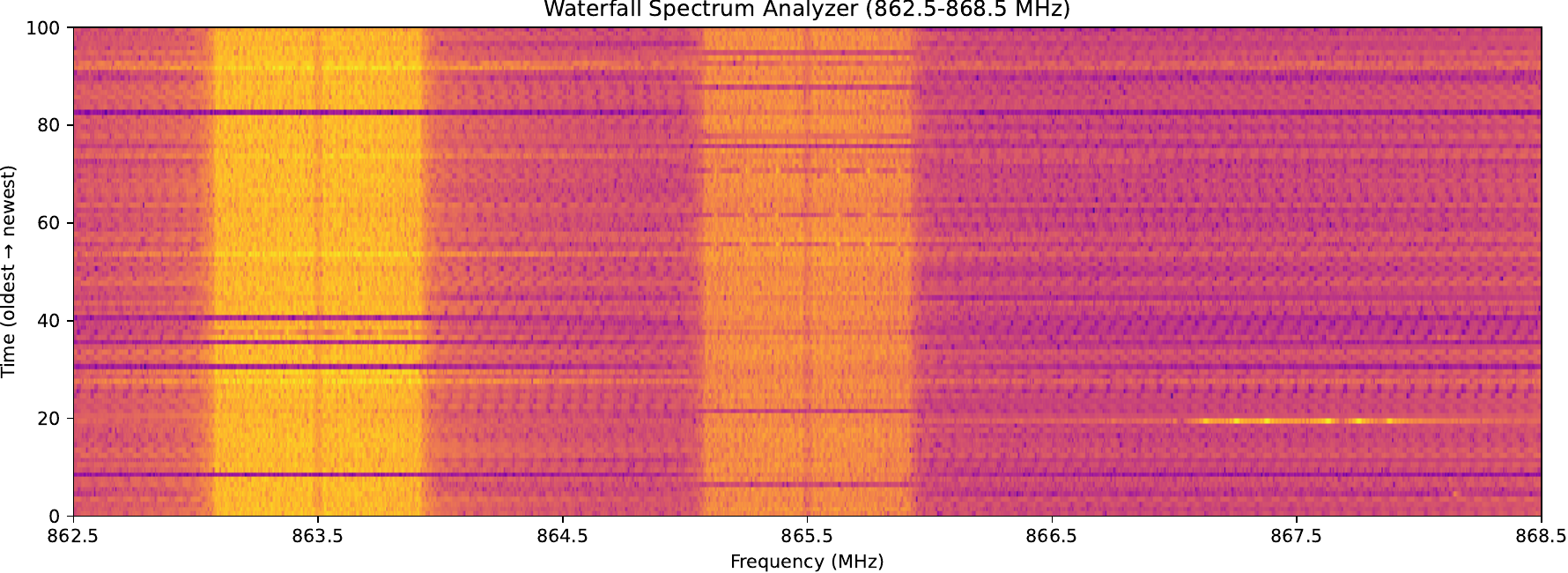}
        \caption{Spectrum Analyzer (Sub-scenario 3)}
        \label{fig:spectrum_analyzer_aci_1_3}
    \end{subfigure}
    
    \caption{Experimental Results - ACI (Section \ref{ACI_2_links})}
    \label{fig:ACI_experimental_results}
    \vspace{-0.45cm}
\end{figure*}

\subsection{ACI with 2 Links} \label{ACI_2_links}
\subsubsection{Description}
Initially, we focus on the most simplistic wireless environment with two pairs of APs and STAs, that interfering with each other. More specifically, five different (\textbf{executed separately}) sub-scenarios are examined, in order to observe the contention and primarily the adjacent channel interference (ACI) occurring between them. We keep "Link 1" (Node5 - Node6) static in $Ch_{1}$ and move the "Link 2" (Node3 - Node4) from $Ch_{1}$ to $Ch_{5}$ gradually at each experimental sub-scenario. This experiment aims to highlight the effect of interference from adjacent channel leakage power, and compared to the free amount of space (frequencies) given between the wireless links each time. 

\subsubsection{Observations}
The results of this experimental scenario are illustrated in Figures \ref{fig:ACI_2links_throughput}, \ref{fig:ACI_2links_packets}. These links, are both closely located to each other and both of them hold excellent wireless AP to STA link conditions. For this reason, it can be observed that both links are performing almost identically, when transmitting separately with no interference (\textit{"Link Capacity without Interference"}). Specifically, 2.35 Mbps and 2.28 Mbps are accomplished from "Link 1" and "Link 2" accordingly. Initially, when both links operate in $Ch_{1}$ the expected performance is achieved from both links, with 0.97 Mbps from "Link 1" and 1.12 Mbps from "Link 2". This is in practice the expected behavior based on Carrier Sense Multiple Access with Collision Avoidance (CSMA/CA) mechanism given in the whole family of IEEE 802.11 standards. The channel's resources seem to be separated identically between the two involved wireless links. 

However, this is not the case at the following executions, when the "Link 2" is being gradually moved away from $Ch_{1}$. In the sub-scenario 2 ("Link 1" = $Ch_{1}$ and "Link 2" = $Ch_{2}$) a performance degradation of nearly 50\% is observed for both links, considering that they already operate in completely non-overlapping channels. Large amount of energy is leaking from neighbor transmissions, enabling Clear Channel Assessment (CCA) mechanism at both links. The Energy Detection (ED) threshold is activated, resulting in larger than expected back-off times and lower resulting performance. This also can be proved from Figure \ref{fig:ACI_2links_packets}, as the succeed / attempted transmission rate of both links are approximately the same, with the sub-scenario in which both links operate at $Ch_{1}$. This is also shown in Figure \ref{fig:spectrum_analyzer_aci_1_2}, where it is obvious that both links are backing off and transmit complementary, at approximately 50\% of the time each one. 

Finally, in the following sub-scenarios 3-5 ("Link 2" gradually operates from $Ch_{3}-_{5}$ at each execution), it can be observed that "Link 1" nearly maximizes its performance, by achieving 2.26 Mbps, 2.32 Mbps and 2.35 Mbps accordingly. In contrary, "Link 2" achieves significantly lower than expected throughput (Figure \ref{fig:ACI_2links_throughput}). More specifically it achieves 1.41 Mbps, 1.67 Mbps and 1.99 Mbps, which are 39\%, 27\% and 13\% lower than expected when operating in completely non-overlapping channels. In contrast with the first two sub-scenarios in which the CSMA/CA and CCA mechanisms seemed to operate properly, at sub-scenarios 3-5 it can be observed from Figure \ref{fig:ACI_2links_packets} a noticeable lower Packet Delivery Rate (PDR) for "Link 2". This highlights that the adjacent channel
leakage power largely affects the performance, but cannot be successfully sensed / avoided through from "Link 2", which results in failed transmissions even in cases that the links operate in distant non-overlapping frequencies (up to 4Mhz). This can be verified indicatively for sub-scenario 3 in Figure \ref{fig:spectrum_analyzer_aci_1_3}, where it is obvious that the links do not back-off and both transmit the whole time. 

Finally, it is noted that all the sub-scenarios were executed in a reverse way, starting both links from $Ch_{5}$ and gradually moving one pair of nodes to $Ch_{1}$. In such a way, we were able verify the same behavior and performance achieved from the two pair of nodes and compared to \ref{fig:ACI_2links_throughput}. Additionally, the energy detected at $Ch_{5}$ in Figures \ref{fig:spectrum_analyzer_aci_1_2},\ref{fig:spectrum_analyzer_aci_1_3}, are due to non-linearity in SDR components, resulting in mixer harmonic outputs of the LO and RF signals.


\begin{figure*}[t!]
    \begin{subfigure}{.47\textwidth}
        \centering
        \includegraphics[width=1\linewidth]{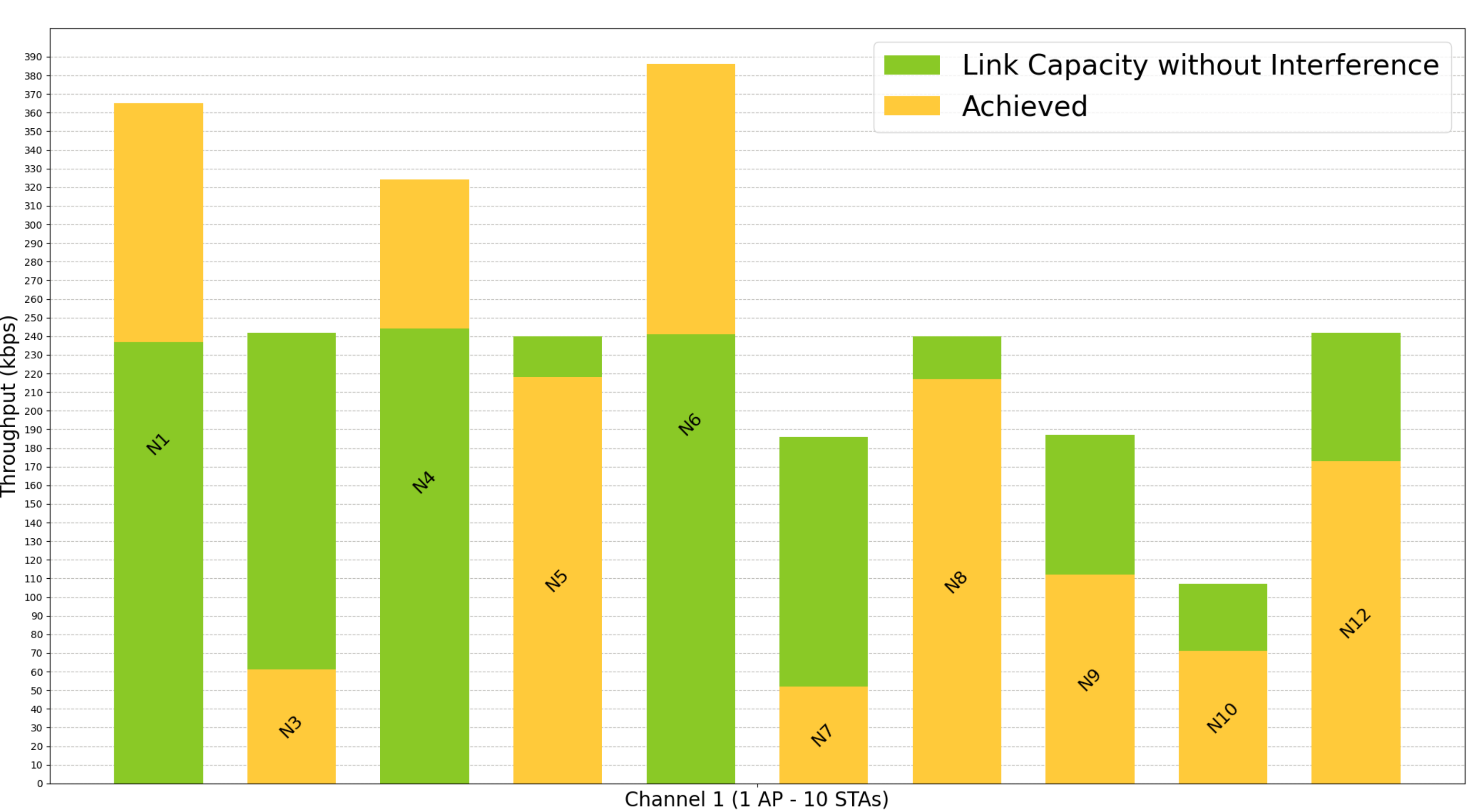}
        \caption{Throughput}
        \label{fig:contention_throughput}
    \end{subfigure}%
    \begin{subfigure}{.469\textwidth}
        \centering
        \includegraphics[width=1\linewidth]{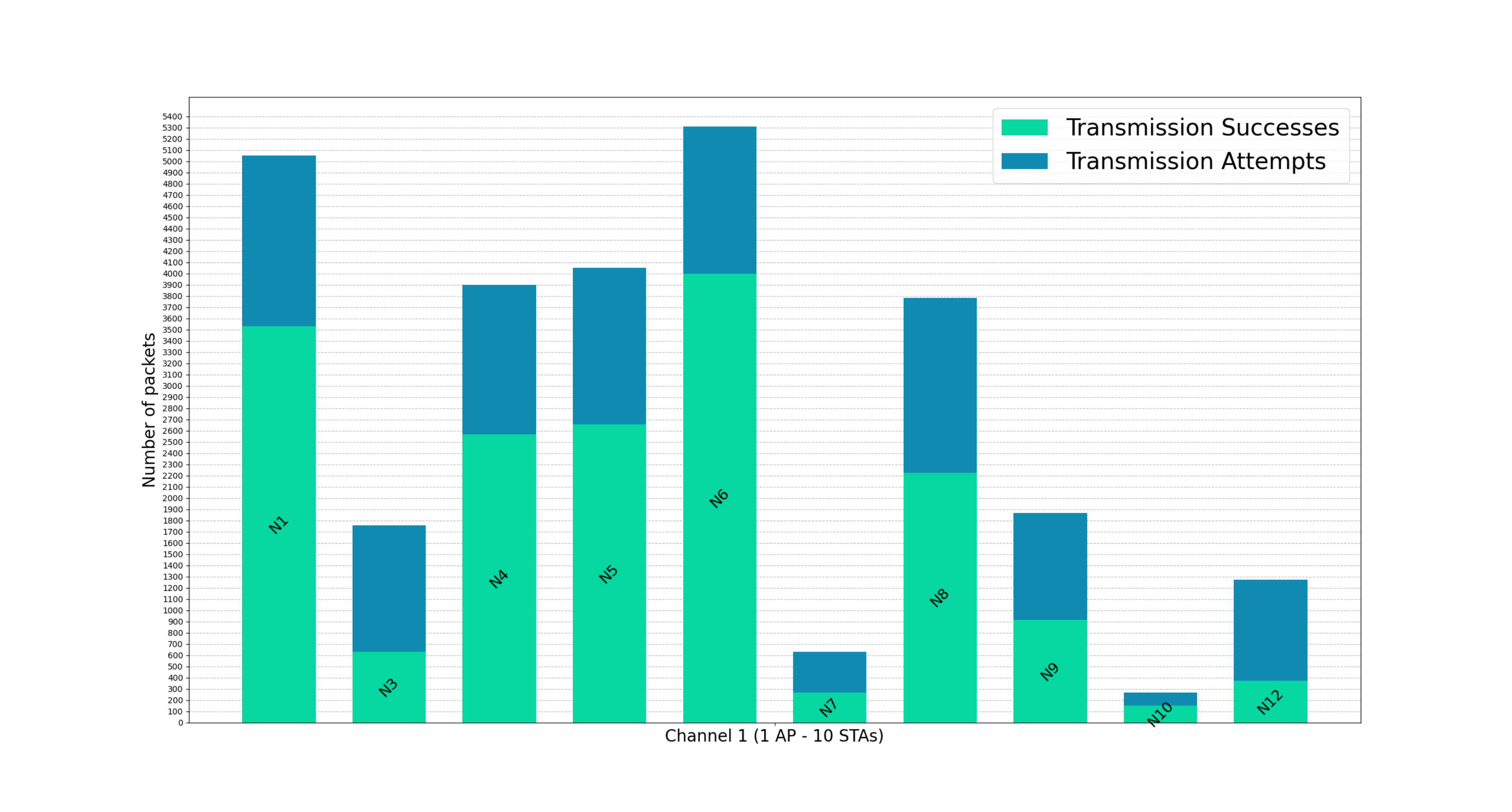}
        \caption{Packet Successes / Attempts}
        \label{fig:contention_packets}
    \end{subfigure}
    
    \caption{Experimental Results - Contention (Section \ref{Contention_10_stations})}
    \label{fig:Contention_experimental_results}
    \vspace{-0.45cm}
\end{figure*}

\begin{figure}[t!]
    \begin{center}
    \includegraphics[width=0.99\linewidth]{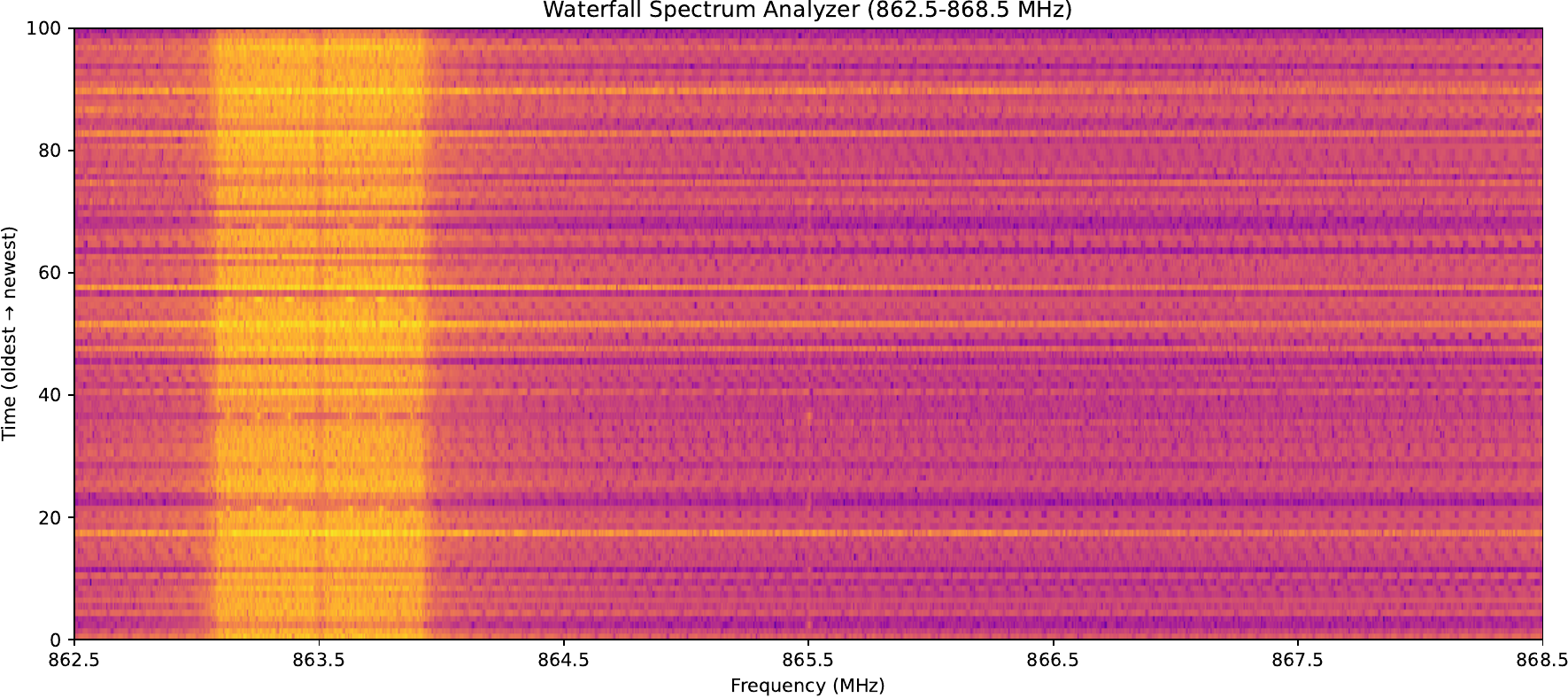}
    \caption{Spectrum Analyzer - Contention (Section \ref{Contention_10_stations})}
    \label{fig:Spectrum_analyzer_contention}
    \vspace{-0.5cm}
    \end{center}
\end{figure}

\subsection{Contention with 10 STAs} \label{Contention_10_stations}

\subsubsection{Description}
In this scenario a single AP (Node 2) with multiple connected STAs (Node 1,3,4,5,6,7,8,9,10,12) set up, is being deployed. The AP operates in $Ch_{1}$ and it is placed at the 3$^{rd}$ floor, while the associated STAs are placed at both floors of the testbed. As the distance between AP and each STA is different, there are also differences on each wireless link quality. This differentiates the MCS selection at each STA, thus resulting in different achievable performance (\textit{"Link Capacity without Interference"}) and transmission attempts at each link. This scenario, aims to examine potential performance losses when intense contention is present in dense IEEE 802.11ah environments.

\subsubsection{Observations}

The results of this experimental scenario are summarized in Figure \ref{fig:Contention_experimental_results}. Specifically, Figure \ref{fig:contention_throughput} contains the \textit{"Link Capacity without Interference"} for each link (kbps), by taking  also into consideration the number of all STAs (10) which are given and contending the same channel. For instance Link (Node 1 - Node 2) actually achieves 2.37 Mbps in clear channel conditions, but 237kbps is noted for that in the graph. This should also be the maximum theoretical achieved throughput in a completely fair channel. However as it is illustrated in Figure \ref{fig:Contention_experimental_results}, this is not the case for real IEEE 802.11ah deployments as the vast majority of the nodes do not achieve the expected performance based on the mechanisms of the IEEE 802.11 standard. In contrary, there are nodes like 1,4 and 6 which achieve more than 1/10th expected. The results indicate the presence of capture effects and potential hidden-terminals caused in the wireless environment, that result in noticeably degraded performance and failed transmissions (collisions) as depicted in Figure \ref{fig:contention_packets}. As expected and also shown in Figure \ref{fig:Spectrum_analyzer_contention}, the spectrum is almost completely occupied over time from the contending STAs during the experiment.


\begin{figure*}[t!]
\centering
\begin{subfigure}{.475\textwidth}
  \centering
  \includegraphics[width=1\linewidth]{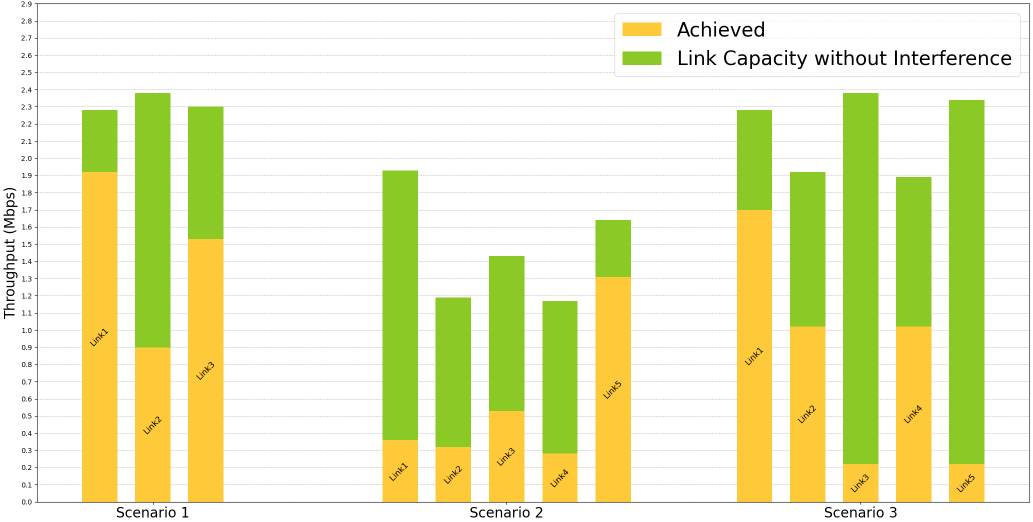}
  \caption{Throughput}
  \label{fig:various_throughput}
\end{subfigure}%
\begin{subfigure}{.47\textwidth}
  \centering
  \includegraphics[width=1\linewidth]{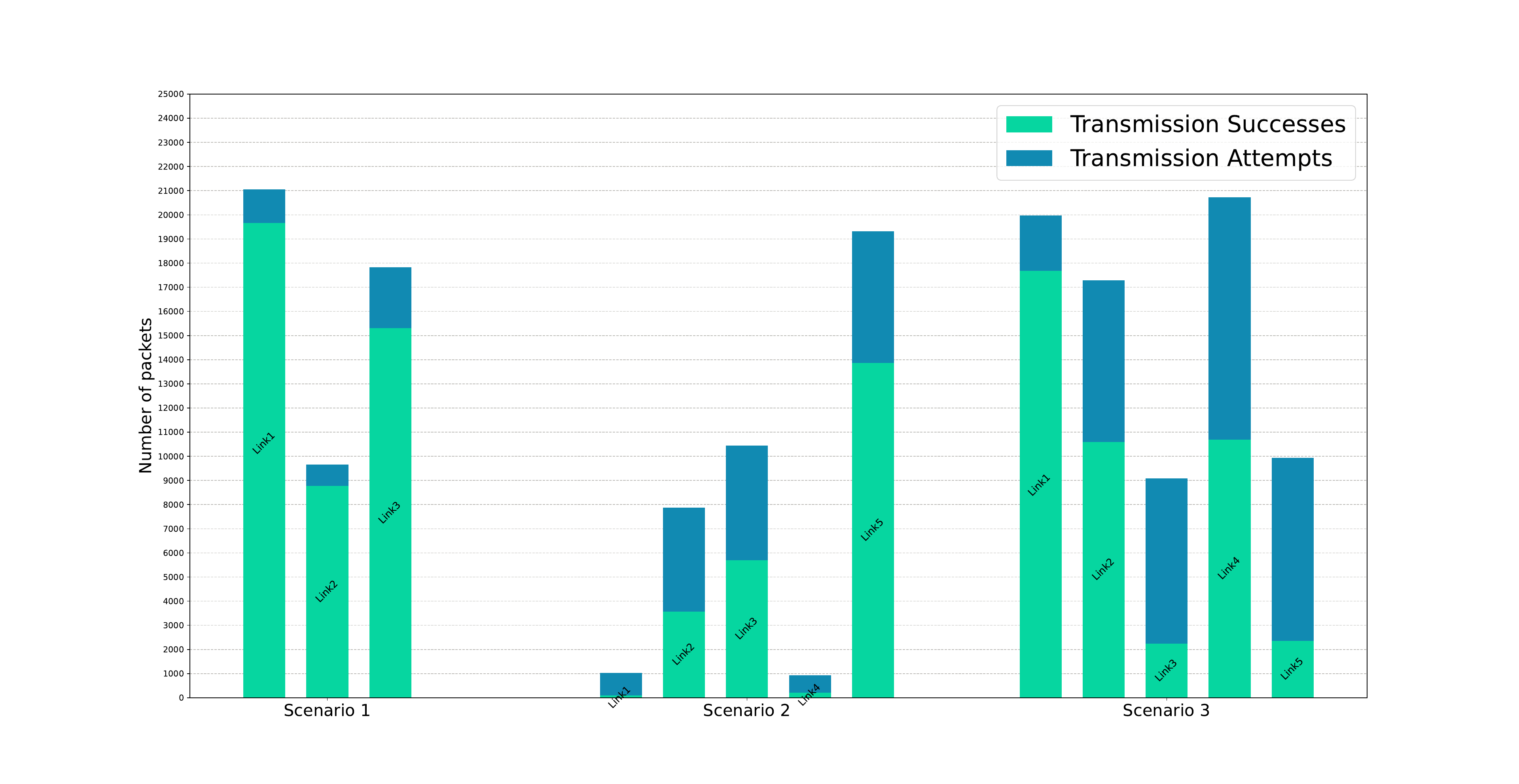}
  \caption{Packet Successes / Attempts}
  \label{fig:various_packets}
\end{subfigure}

\caption{Experimental Results - Various Scenarios (Section \ref{various_scenarios})}

\label{fig:experimental_results_various}
\vspace{-0.45cm}
\end{figure*}

\begin{figure}[t!]
    \begin{center}
    \includegraphics[width=0.99\linewidth]{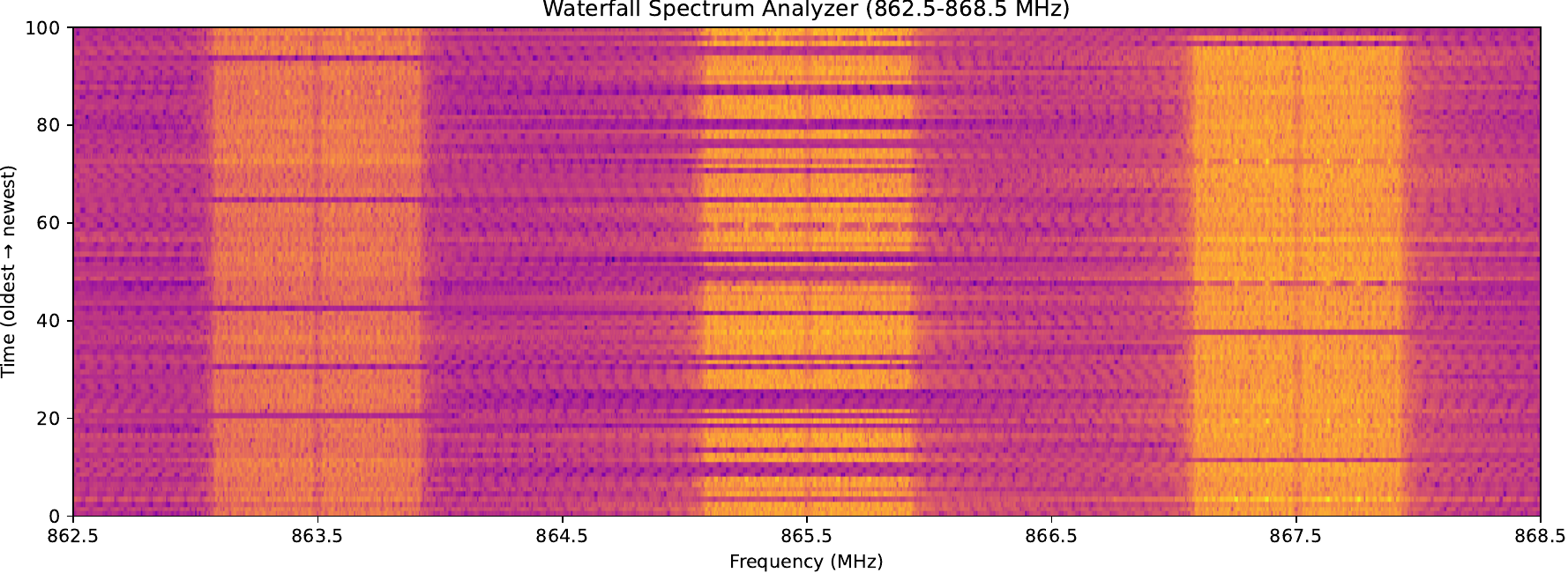}
    \caption{Spectrum Analyzer - Var. - Scenario 1 (Section \ref{various_scenarios})}
    \label{fig:Spectrum_analyzer_various_scenario_1}
    \vspace{-0.5cm}
    \end{center}
\end{figure}

\subsection{Various Scenarios ($\geq3$ links)} \label{various_scenarios}
Finally, additional scenarios were executed in the developed indoor testbed, for examining the performance achieved, when three or more links, are transmitting simultaneously in different channels. These experiments highlight that the performance further deteriorates when more links are deployed in complex dense environments. The experimental results for those three scenarios are depicted in Figure \ref{fig:experimental_results_various}.   

\subsubsection{Observations}


In Scenario 1, we deploy three wireless links, that are all located on the 3$^{rd}$ floor of the laboratory. More specifically, "Link 1" (Node5 - Node6) Link 2" (Node1 - Node2) and "Link 3" (Node3 - Node4), operating in $Ch_{1}$, $Ch_{3}$ and $Ch_{5}$ accordingly. Based on the channel selection scheme, it is noted that the three links are equally distributed in the available spectrum. However, as it is depicted in Figures \ref{fig:various_throughput}, \ref{fig:various_packets} noticeable performance degradation is observed for all links. More specifically, 1.92 Mbps ($\sim -16\%$), 0.89 Mbps ($\sim -62\%$) and 1.53 Mbps ($\sim -33\%$) are noted from the involved wireless links, compared to the 2.28 Mbps, 2.38 Mbps and 2.3 Mbps, that can be achieved on clear channel conditions for each of them. Moreover, it is obvious from Figure \ref{fig:Spectrum_analyzer_various_scenario_1}, that in the case of Link 2 the ED threshold is exceeded by the ongoing $Ch_{1}$ and $Ch_{5}$ transmissions, resulting in larger than expected back-off compared to Link 1 and Link 3. On the one hand, this results in lower performance achieved, but on the other hand with fairly high PDR for Link 2, as depicted in Figure \ref{fig:various_packets}.    


In Scenario 2 the number of links is further increased from three to five. Additionally, this is the first scenario in which there are links with nodes combined from both 2$^{nd}$ and 3$^{rd}$ floors. Specifically, the deployed links are "Link 1" (Node2 - Node8), "Link 2" (Node1 - Node7), "Link 3" (Node3 - Node9), "Link 4" (Node4 - Node10) "Link 5" (Node6 - Node12) and they operate from $Ch_{1}$ to $Ch_{5}$ accordingly. This greater distance between APs and STAs, leads to the slightly lower link capacity without interference for all the involved links as illustrated in Figure \ref{fig:various_throughput}. In this scenario even worse performance is observed, with "Links 1-4" to be able to perform at 0.36 Mbps, 0.32 Mbps, 0.53 Mbps and 0.28 Mbps.  The expected capacities for these links were 1.93 Mbps, 1.19 Mbps, 1.43 Mbps and 1.17 Mbps, noticing there a link performance degradation of -82\%, -74\%, -63\% and -77\%. 

For Scenario 3, we follow a similar approach to Scenario 2 by utilizing nodes both from the 2$^{nd}$ and 3$^{rd}$ floors. However, in this case each link is made up solely from nodes of the same floor. Analytically, the links involved in this scenario are "Link 1" (Node1 - Node2), "Link 2" (Node7 - Node8), "Link 3" (Node5 - Node6), "Link 4" (Node11 - Node12) and "Link 5" (Node3 - Node4) and they operate from $Ch_{1}$ to $Ch_{5}$ accordingly. In this scenario, a slightly better overall performance it is observed for the majority of the links, when compared to Scenario 2. However, this is not the case for "Links 3 \& 5" which achieve 0.22 Mbps and 0.23 Mbps. These links achieve only $\sim 10\%$ of their overall expected capacity.  

\subsection{Energy Consumption Considerations} \label{energy_consumption}

Energy consumption reduction is crucial across the whole IoT-to-cloud continuum, where in many cases, the resource-constrained edge devices continuously interact with high-performance cloud infrastructures. Nowadays, there is an increased number of initiatives that focus on lowering the environmental impact and operation costs on research infrastructures. In the context of this work, we revealed several scenarios / topologies in the developed IoT testbed, in which the performance achieved is unexpectedly reduced. Lower throughput leads to longer transmission and processing times, causing devices to remain active for extended periods. This prolonged activity results in higher overall energy consumption, as wireless transmissions and computational tasks consume significant power when sustained over time. Thus, we have indicatively measured the power consumption on a testbed node for idle, transmit, and receive states. 

\begin{figure}[t!]
    \begin{center}
    \includegraphics[width=0.99\linewidth]{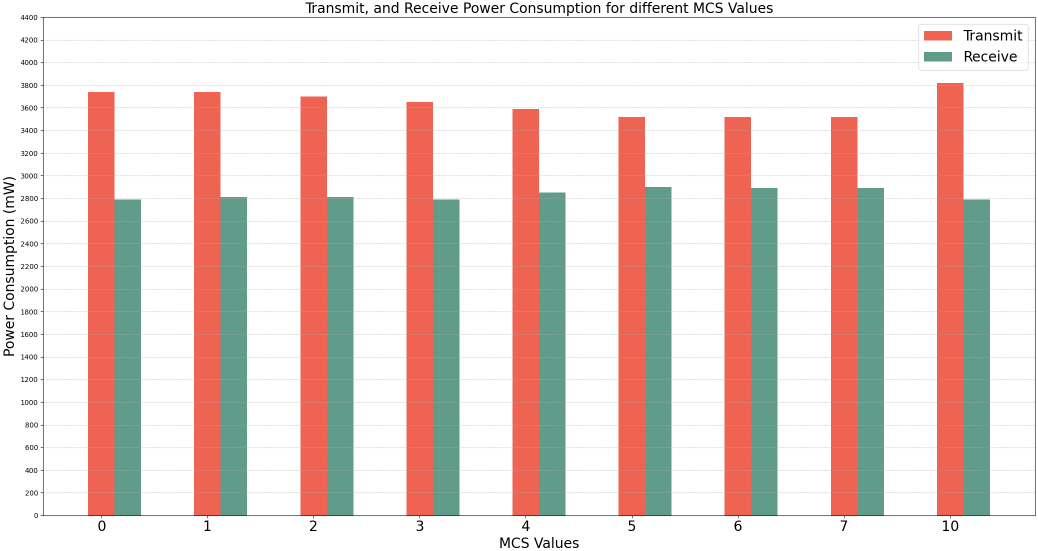}
    \caption{Energy Consumption per MCS}
    \label{fig:Energy_consumption_per_MCS}
    \end{center}
\end{figure}

Figure \ref{fig:Energy_consumption_per_MCS} shows the IEEE 802.11ah adapter's energy consumption, for each available MCS. Initially, to measure the energy consumed by the node, we utilized Shelly Plug S \cite{Shelly}, which is capable of providing energy measurements with 0.01 W accuracy. A Python script was also developed which collects power measurements every 0.5 seconds through the Shelly Plug API, store them, and exports the average value at the end of the experiment. More specifically, after obtaining power measurements for 120 seconds, while the node was in an idle mode (not transmitting / receiving), 2750 mW was extracted as the node's average power consumption. This value includes the standard operating system's processes and services running on each testbed node. Thereafter, saturated UDP traffic was initiated to calculate the energy consumed during the receive and transmit states. The results depicted in Figure \ref{fig:Energy_consumption_per_MCS}, include idle energy consumption, IEEE 802.11ah adapter consumption, and any additional impact arising from generating and encoding/decoding network packets. 

It should be noted that IEEE 802.11ah utilizes MCS0 (BPSK) as the basis, and reaches gradually up to more complex modulation and coding schemes like MCS7 (64-QAM). The protocol also supports MCS8-9 (256-QAM), but these schemes are not yet achievable on the commercial hardware available. Furthermore, MCS10 usually used for extended range, and robustness in low-SNR environments, as it holds again a BPSK modulation and an even more simple 1/2 x 2 coding rate, compared to this given in MCS0. 
The values observed when the node received UDP traffic were between 40 - 150 mW, with a marginal increase from MCS0 to MCS7, most likely because the node processes more data at higher MCS selections. Furthermore, 770 - 1070 mW were consumed during the transmission of saturated UDP data from the wireless adapter, noting in parallel that the higher power consumption is reduced in higher MCS. This is absolutely normal, taking into consideration the shorter transmission times (airtime) of the higher modulation and coding schemes. Finally, all the measurements  noted for the transmit / receive energy consumption of this IEEE 802.11ah wireless adapter, were captured using an AP / STA link, and by ensuring clear channel conditions with the deployed spectrum analyzer.


\section{Conclusions} \label{conclusions}

In this work, an IEEE 802.11ah testbed infrastructure has been developed, where several challenging experimental scenarios have been thoroughly tested. Specifically, the impact of Adjacent Channel Interference (ACI) and heavy contention has been examined by deploying a high density of IEEE 802.11ah links within a constrained space. The findings of this work highlight that abnormal performance degradation may occur, in some cases reaching up to $\sim 90\%$ for the wireless links under consideration. This study not only measures the impact of unexpected interference in isolated non-overlapping channels of the IoT sub-GHz unlicensed band, but also lays the groundwork for an automated ML-driven mechanism which already considered as a future work. The foreseen mechanism will aim to dynamically detect and mitigate similar spectrum pathologies by intelligently adjusting transmission parameters at the device level. This will result in severe improvements both at network performance and energy consumption in real-time.
\section{Acknowledgments}
The research leading to these results has received funding from the European Horizon Europe Research and innovation funding programme under Grant Agreement Number No 101131207 (Horizon Europe GreenDIGIT).The European Union and its agencies are not liable or otherwise responsible for the contents of this document; its content reflects the view of its authors only..

\bibliographystyle{IEEEtran}
\bibliography{literature}

\end{document}